%% file: draft-v6.tex
\documentclass[twocolumn]{aastex61}
\usepackage{natbib}
\usepackage{verbatim}
\usepackage{amsmath}
\usepackage{gensymb}
\usepackage{graphicx}

\newcommand{\bdv}[1]{\mbox{\boldmath$#1$}}

\def\au{{\rm AU}}

\def\kms{{\rm km}\,{\rm s}^{-1}}
\def\masyr{{\rm mas}\,{\rm yr}^{-1}}
\def\kpc{{\rm kpc}}

\def\muas{\mu{\rm \,as}}
\def\rel{{\rm rel}}
\def\hel{{\rm hel}}

\def\e{{\rm E}}
\def\bpi{{\bdv\pi}}
\def\bmu{{\bdv\mu}}

\begin{document}
\title{OGLE-2017-BLG-1130: THE FIRST BINARY GRAVITATIONAL MICROLENS DETECTED FROM {\it Spitzer} ONLY}
\input{author.tex}

\begin{abstract}
We analyze the binary gravitational microlensing event OGLE-2017-BLG-1130 (mass ratio $q \sim0.45$), the first published case in which the binary anomaly was only detected by the {\it Spitzer} Space Telescope. This event provides strong evidence that some binary signals can be missed by observations from the ground alone but detected by {\it Spitzer}. We therefore invert the normal procedure, first finding the lens parameters by fitting the space-based data and then measuring the microlensing parallax using ground-based observations. We also show that the normal four-fold space-based degeneracy in the single-lens case can become a weak eight-fold degeneracy in binary-lens events. Although this degeneracy is resolved in event OGLE-2017-BLG-1130, it might persist in other events. 
\end{abstract}

\section{Introduction}\label{sec:intro}
\indent\indent The projected Einstein radii, $\tilde r_\e\equiv {\rm AU}/\pi_\e$, of typical Galactic gravitational microlensing events are of the order of a few astronomical units (au). Hence, the relative lens-source positions seen from the ground and from a satellite in solar orbit appear to be different. This results in different light curves, and a combined analysis of the light curves should lead to the measurement of the microlens-parallax vector $\bdv{\pi_{\rm E}}$ \citep{Refsdal:1966,Gould:1994}, which is related to the physical lens parameters by 
\begin{equation}
\bdv{\pi_{\rm E}}\equiv\frac{\pi_{\rm rel}}{\theta_{\rm E}}\frac{\bdv{\mu}}{\mu};\quad \theta_{\rm E}\equiv \sqrt{\kappa M\pi_{\rm rel}}
\label{piedef}
\end{equation} 
where $(\pi_{\rm rel},\bdv{\mu})$ are the lens-source relative (parallax, proper motion), $\theta_E$ is the angular Einstein radius, $\kappa \equiv 4G/(c^2{\rm au})\simeq8.14\ {\rm mas}/M_{\odot}$.
The measurement of $\bdv{\pi_{\rm E}}$ is important because, by itself, it strongly constrains the lens mass $M$ and distance $D_L$ \citep{Han:1995}, and provided that $\theta_\e$ is also determined, enables one to measure both quantities,
\begin{equation}\label{equ:mass}
M = \frac{\theta_E}{\kappa \pi_{\rm E}};\quad D_L = \frac{\rm AU}{\pi_{\rm E}\theta_E+\pi_S},
\end{equation}
where $\pi_S={\rm AU}/D_S$ is the parallax of the lensed star (source), and $D_S$ is the distance to the source

Since 2014, {\it Spitzer} Space Telescope has measured the microlens parallax $\bdv{\pi_{\rm E}}$ for hundreds of microlensing (single and binary) events, proving it to be an excellent microlensing parallax measurement satellite \citep{Dong:2007,ob140124,Yee15,Zhu:2015,Novati:2015}. However, the usefulness of {\it Spitzer} observations in characterizing binaries can extend beyond measuring parallax. For example, in the case of event OGLE-2014-BLG-0124 \citep{ob140124}, the planetary signal was independently detected from {\it Spitzer}.

For most binary microlensing events, the caustic structures and lens parameters can be directly determined by ground-based observations. In such cases, {\it Spitzer} data are only used to measure the satellite parallax and sometimes to resolve the remaining degeneracies. Nevertheless, it is possible, in principle, that the binary signal would be detected solely from the satellite, in which case the ground-based data would be used to measure the parallax parameters. This happens in the binary event OGLE-2017-BLG-1130, for which the ground-based data show no deviation from a single-lens light curve. 

Light curves of single-mass lensing events obtained from ground-based observatories and one space-based observatory typically yield a set of four degenerate solutions \citep{Refsdal:1966,Gould:1994,Gould&Horne2013,Novati:2015,Yee:2015},
\footnote{This four-fold degeneracy can be resolved if observations from a second satellite are obtained, as pointed out by \citet{Refsdal:1966} and \citet{Gould:1994} and recently demonstrated by \citet{Zhu:2017b}.}
which are often denoted as (+,+), ($-$,$-$), (+,$-$), and ($-$,+). Here the first and second signs in each parenthesis represent the signs of the lens-source impact parameters as seen from Earth and from the satellite, respectively. See Figure~4 of \citet{Gould:2004} for the sign conventions. This four-fold degeneracy can be expressed as (+,$-$)$\times$(same, opposite), where the signs in the first parenthesis represent the signs of the impact parameter seen from Earth, and same (or opposite) means that the source trajectories seen from Earth and from the satellite pass on the same side (or opposite sides) with respect to the lens.	

For binary-lens events that are well covered by the observations, the (same, opposite) degeneracy is generally broken due to the asymmetry in the light curve. If the degeneracy remains unresolved, we usually consider that ``opposite side of the lens'' means ``opposite side of the nearby component of lens primary'', as has been seen in previous cases \citep{Zhu:2015,Han:2016}. For binary-lens events that are not well covered, such as the present event OGLE-2017-BLG-1130, the source trajectories seen from Earth and the satellite in the ``opposite'' solution can in principle pass on the opposite side of either the whole lens system or of the nearby component (see Section~\ref{4fold}), and there can be as many as eight degenerate solutions. We identify this new form of the four-fold degeneracy here for the first time. We show that while it is resolved for OGLE-2017-BLG-1130, it may persist in the case of other events.

In this paper, we present the analysis of the {\it Spitzer} binary event OGLE-2017-BLG-1130. This is the first published case in which the binary anomaly is detected by {\it Spitzer} only. We summarize the ground-based and space-based observations in Section~\ref{sec:obs}, describe the light curve modeling in Section~\ref{sec:model}, and derive the physical properties of the binary system in Section~\ref{sec:phys}. In Section~\ref{sec:dis}, we discuss the potentially new form of the four-fold degeneracy that occurs in this event.

\section{Observations}\label{sec:obs}
\subsection{Ground-Based Alert and Follow-up}
\indent \indent At UT 11:57 of 2017 June 19 (HJD$'$ = HJD$-$2450000 = 7924.00), the OGLE collaboration identified the microlensing event OGLE-2017-BLG-1130 at equatorial coordinates (R.A., decl.)$_{2000}$ = ($18^{\rm h}01^{\rm m}36\fs93$, $-27\arcdeg39\arcmin56\farcs9$) with corresponding Galactic coordinates $(l,b)$ = (2.88$^\circ$,  $-$2.39$^\circ$), based on observations with the 1.3m Warsaw telescope with 1.4 {\rm deg}$^2$ camera at Las Campanas in Chile. This microlensing event, lies in the OGLE-IV field BLG511, which was covered with a cadence of $1\,{\rm hr^{-1}}$ \citep{ews1,ews2, ews3}.

The Korea Microlesning Telescope Network (KMTNet, \citealt{kmtnet}) observed this event from its three 1.6m telescopes at CTIO (Chile, KMTC), SAAO (South Africa, KMTS) and SSO  (Australia, KMTA), in its BLG03 field, with cadence of $2\,{\rm hr^{-1}}$. KMTNet designated the event as BLG03K0102.032555.

All ground-based data were reduced using variants of the image subtraction method \citep{Alard:1998,wozniak2000,albrow09}.

Both OGLE and KMTNet data were adversely affected by a diffraction spike from a nearby bright star.  Because this star was very blue, the $V$-band light curves from both surveys were
completely corrupted.  Since these would normally be used to determine the source color $(V-I)_S$, we had to develop a novel technique to measure this quantity, a point to which we return below.

The $I$-band light curves also suffered from some degradation depending on the observatories (OGLE, KMTS, KMTC, KMTA) where the data were taken. Because the ground-based data are well-characterized by a \citet{pac86} ``point lens'' fit, we could afford to be quite conservative in including in the modeling only the best ground-based data.   We found that the OGLE and KMTS data were of comparable, and generally quite good, quality. On the other hand, the KMTC and KMTA data showed much larger scatter and also much greater systematics.  We therefore do not use KMTC and KMTA data in our analysis.  Closer investigations of the OGLE and KMTS data revealed that both display some systematics in ``better seeing'' images.  This is not surprising because diffraction spikes are more pronounced in better seeing.  Although these effects were not severe, to be conservative, we nevertheless eliminated all OGLE images with FWHM$<1.17^{\prime\prime}$ (4.5 pixels) and all KMTS images with FWHM$<2.08^{\prime\prime}$ (5.2 pixels).

\subsection{{\it Spitzer} Follow-up}
OGLE-2017-BLG-1130 was originally selected as a {\it Spitzer} target within the framework of the protocols of \citet{yee15}.  These protocols are designed to obtain an ``objective sample'' to measure the Galactic distribution of planets despite the fact that humans must make observing decisions based on real-time data.  Very briefly, events can be selected ``objectively'', ``subjectively'', or ``secretly''.  Events that meet certain objective criteria {\it must} be observed according to the pre-specified rules.  As a consequence, all planets found in the data enter into the Galactic-distribution sample. Events can be selected subjectively by the {\it Spitzer} team for any reason.  However, only planets that do not give rise to significant signal in the data available at the time of the announcement can be included in the Galactic-distribution sample.  The announcement must specify the candence of observations and the time (or conditions) under which the observations will cease.  Finally, events can be selected ``secretly'', i.e., without public announcement.  In this case {\it Spitzer} observations are commenced with no specifications on when they might terminate.  Such events may be converted by the team from ``secret'' to ``subjective'' by making a public announcement.  In this case, planets can enter the Galactic-sample according to the conditions governing ``subjective'' events, and in particular, according to the date of the public announcement.

OGLE-2017-BLG-1130 was initially chosen ``secretly'' on June 19, just a few hours after it was announced by OGLE (and just before the {\it Spitzer} upload time) because it was judged by the upload subteam that it might reach relatively high magnification based on the data then available.  In particular, this subteam does not generally have the authority to choose events subjectively without consulting the team, other than in exceptional circumstances.  The following week, the event's future course remained too uncertain to decide between stopping observations and choosing it subjectively. Hence, it remained ``secret''.  Finally, at UT 16:56 on July 2, shortly before the third upload, it was publicly announced as ``subjective''.  Observations
continued until the end of the {\it Spitzer} window. The binary signal in the {\it Spitzer} data only became discernible at UT 00:28 on August 9, i.e., five days after the final observation, when the reduced {\it Spitzer} data were circulated to the team.  It was specifically noted by SCN about 16 hours later. Because OGLE-2017-BLG-1130L is not planetary, these details do not directly impact any scientific conclusion.  However, we document them here nonetheless in order to maintain homogeneous records for planets and binaries.

\section{Light Curve Modeling}\label{sec:model}
\subsection{Initial Solution Search}
We fit a binary microlensing model to the light curve to explain the observed variation in brightness. The standard binary modeling needs seven basic parameters: the time of the source closest approach to the center of mass of the lens system, $t_0$; the impact parameter with respect to the center of mass of the lens system normalized by the Einstein radius, $u_0$; the Einstein radius crossing time, $t_\e\equiv\theta_{\rm E}/\mu$, where $\mu$ is the relative lens-source proper motion; the source radius normalized by the Einstein radius, $\rho\equiv\theta_\star/\theta_{\rm E}$; the projected separation of the binary components normalized to the Einstein radius, $s$; the binary mass ratio, $q$; and the angle between the binary-lens axis and the lens-source relative motion, $\alpha$. With these seven parameters, we can calculate the binary magnification as a function of time $A(t)$. To describe the blend in the crowded stellar fields, we further introduce two flux parameters, the source flux $(F_{\rm S,j})$ and the blending flux $(F_{\rm B,j})$ so that the observed flux at given time $t_i$ is
\begin{equation}
    F_{\rm j}(t_i) = F_{\rm S,j} \cdot A(t_i)+F_{\rm B,j}\ .
\end{equation}
where $A(t_i)$ is the magnification at $t_i$. These flux parameters are found for each data set and each trial of geometric parameters from a linear fit.

We calculate the binary lens magnification $A(t)$ using the advanced contour integration code, \texttt{VBBinaryLensing}
\footnote{\url{http://www.fisica.unisa.it/GravitationAstrophysics/VBBinaryLensing.htm}}. This code includes a parabolic correction in Green's line integral that automatically adjusts the step size of integration based on the distance to the binary caustic, in order to achieve a desired precision in magnification. See \citet{Bozza:2010} for more details.

\begin{figure*}
\centering
\centering
\includegraphics[width=0.48\textwidth]{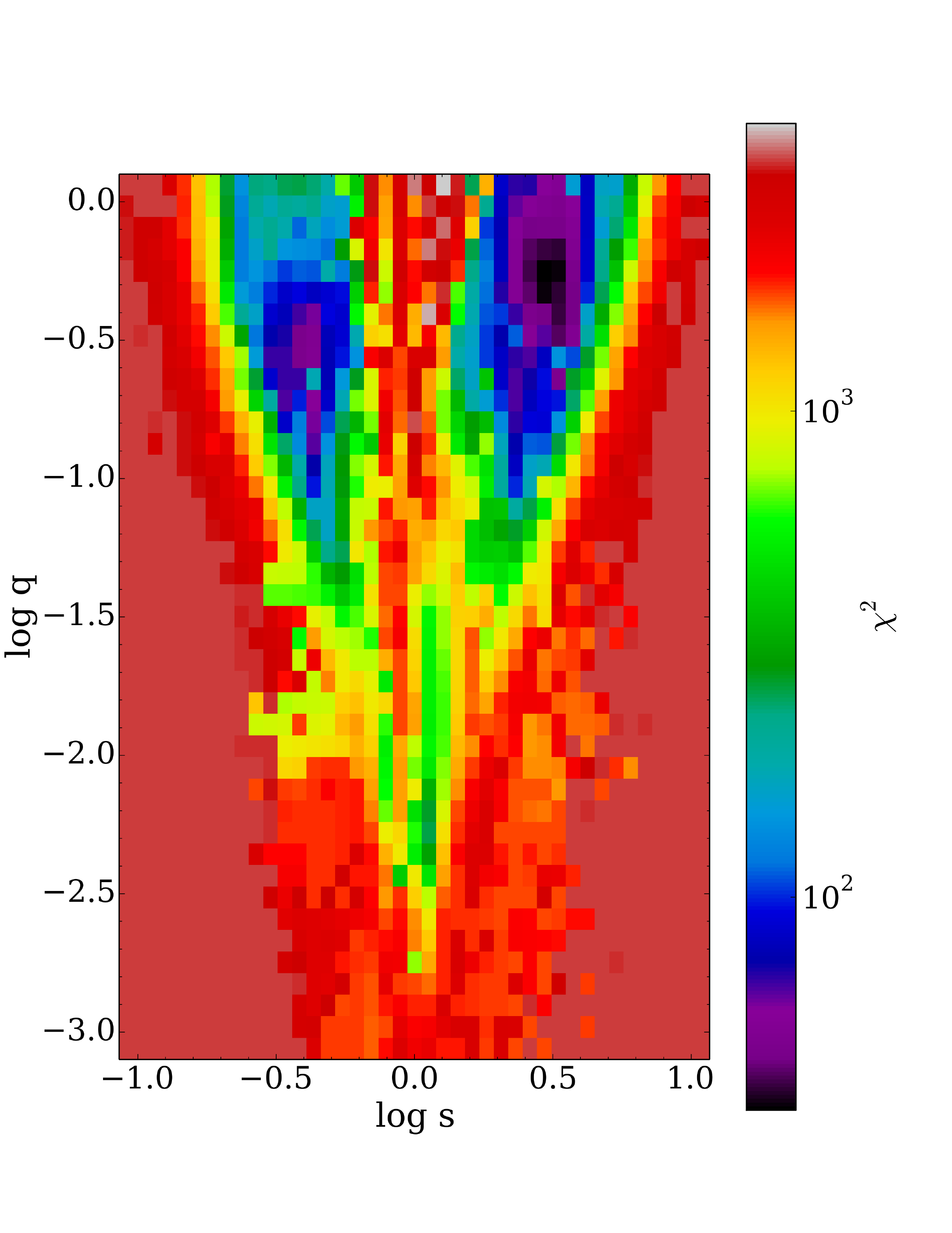}
\caption{$\chi^2$ results of the grid search projected onto the ($\log s$, $\log q$) plane.} \label{fig:grid}
\end{figure*}

To find the best-fit model, we first fit the {\it Spitzer} data only, since the binary signal is not detected from ground. We conducted a grid in the ($\log s$, $\log q$, $\alpha$) parameter space, with 40 values equally spaced between $-1 \leq$ $\log s$ $\leq1$, $-3 \leq$ $\log q$ $\leq0$ and $0^\circ$$ \leq \alpha \leq 360^\circ$, respectively. For each set of ($\log s$, $\log q$, $\alpha$), we find the minimum $\chi^2$ by using a function based on the Nelder-Mead simplex algorithm from the SciPy package\footnote{See \url{https://docs.scipy.org/doc/scipy/reference/generated/scipy.optimize.fmin.html\#scipy.optimize.fmin}.} on the remaining parameters ($t_0$, $u_0$, $t_E$, log $\rho$). We find the global minimum at $\log s \sim 0.5$, $\log q \sim -0.3$ and $\alpha \sim 108^\circ$, and the result of the grid search clearly shows the close-wide degeneracy (See Figure~\ref{fig:grid}). Other local minima will be discussed in Section~\ref{minima}. 

We then perform a Markov Chain Monte Carlo (MCMC) analysis on all parameters around the initial solutions found by the previous grid search, which employs the \texttt{emcee} ensemble sampler \citep{ForemanMackey:2013}. 

\subsection{Inclusion of the Microlensing Parallax Effect}
\indent \indent The microlensing parallax effect must be taken into account in order to simultaneously model the ground-based and space-based data. This effect invokes two additional parameters, $\pi_{\rm E,N}$ and $\pi_{\rm E,E}$, which are the northern and eastern components of the microlens parallax vector $\bdv{\pi_{\e}}$. We extract the geocentric locations of {\it Spitzer} during the entire season from the \emph{JPL Horizons} website
\footnote{\url{http://ssd.jpl.nasa.gov/?horizons}}
 and project them onto the observer plane. The projected locations are then oriented and rescaled according to a given $\bdv{\pi_{\rm E}}$ to determine the lens-source vector as seen from  {\it Spitzer}.

As described in Section~\ref{sec:intro}, the normal four-fold space-based degeneracy \citep{Refsdal:1966, Gould:1994} for single lens events $[(+,+),(-,-),(+,-),(-,+)]$ is potentially ambiguous when applied to binary events.  That is, for single-lens events, this degeneracy can be expressed as $(+,-)\times$(same, opposite), where ``same'' and ``opposite'' refer to the location of the source trajectory as seen from the satellite relative to the location as seen from Earth.  However, for a wide binary lens, this degeneracy can become six fold: $(+,-)\times$(same, opposite nearby component, opposite whole binary). In some cases, the ``opposite nearby component'' will become two-fold degenerate with the trajectory closer to the primary star or to the secondary star (See Section~\ref{4fold}). Therefore, in the most general case, there would be eight degenerate solutions. This form of degeneracy was not previously anticipated and appears for the case of OGLE-2017-BLG-1130 for the first time. The parameters of these solutions are shown in Table~\ref{parms1}.

\subsection{Summary of Local Minima}\label{minima}
For completeness, we present all local minima in this section. These minima can all explain the data qualitatively. As we subsequently show, however, only the pair of $(+,-)\times$(same) solutions, i.e., the $(+,+)$ and $(-,-)$ solutions, are viable.   

\subsubsection{Best-fit Model}\label{4fold}
\begin{figure}
\centering
\includegraphics[width=1.0\textwidth]{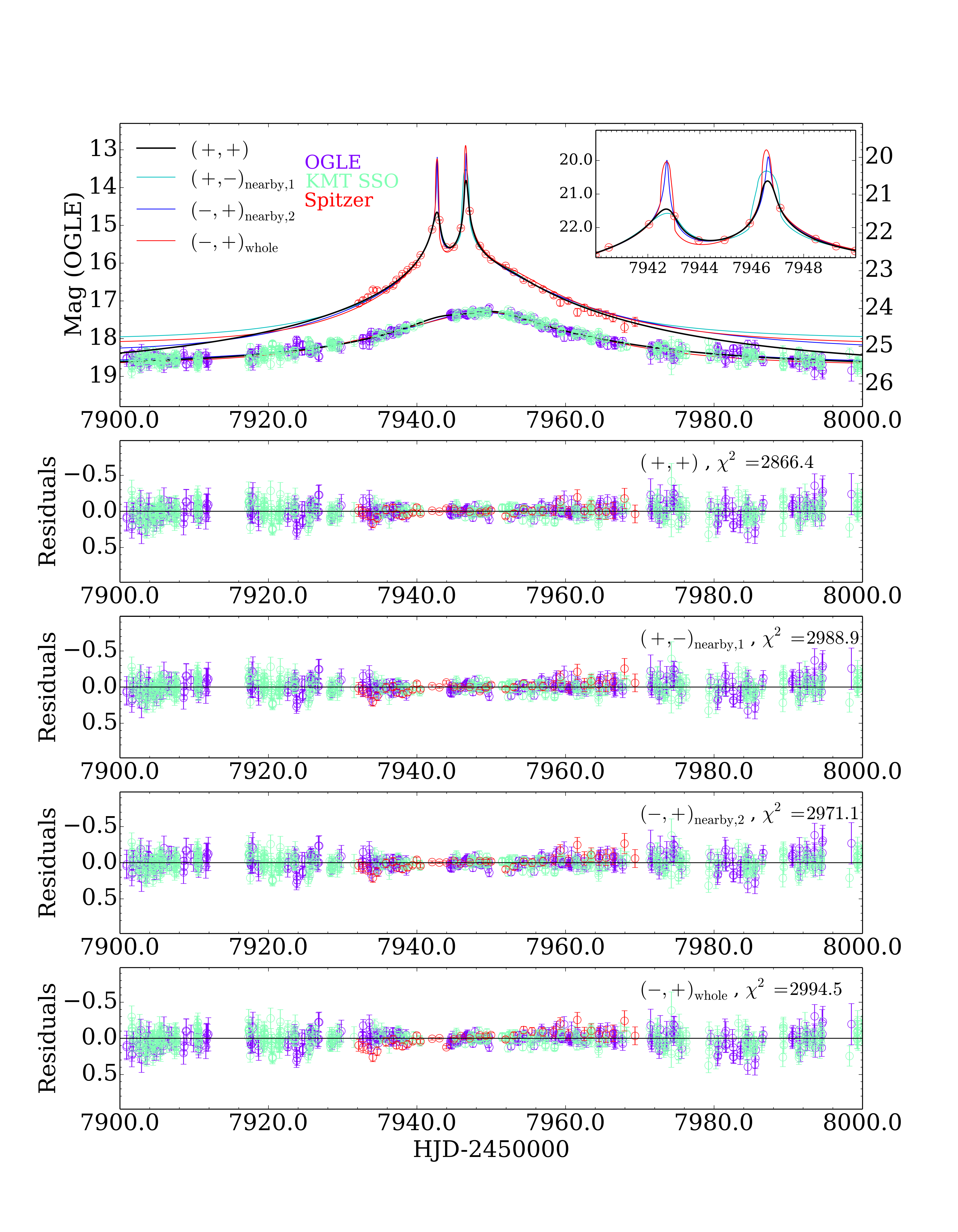}
\caption{Light curves of the best-fit model and its degenerate counterparts.}\label{lc}
\end{figure}

\begin{figure}
\centering
\includegraphics[width=0.48\textwidth]{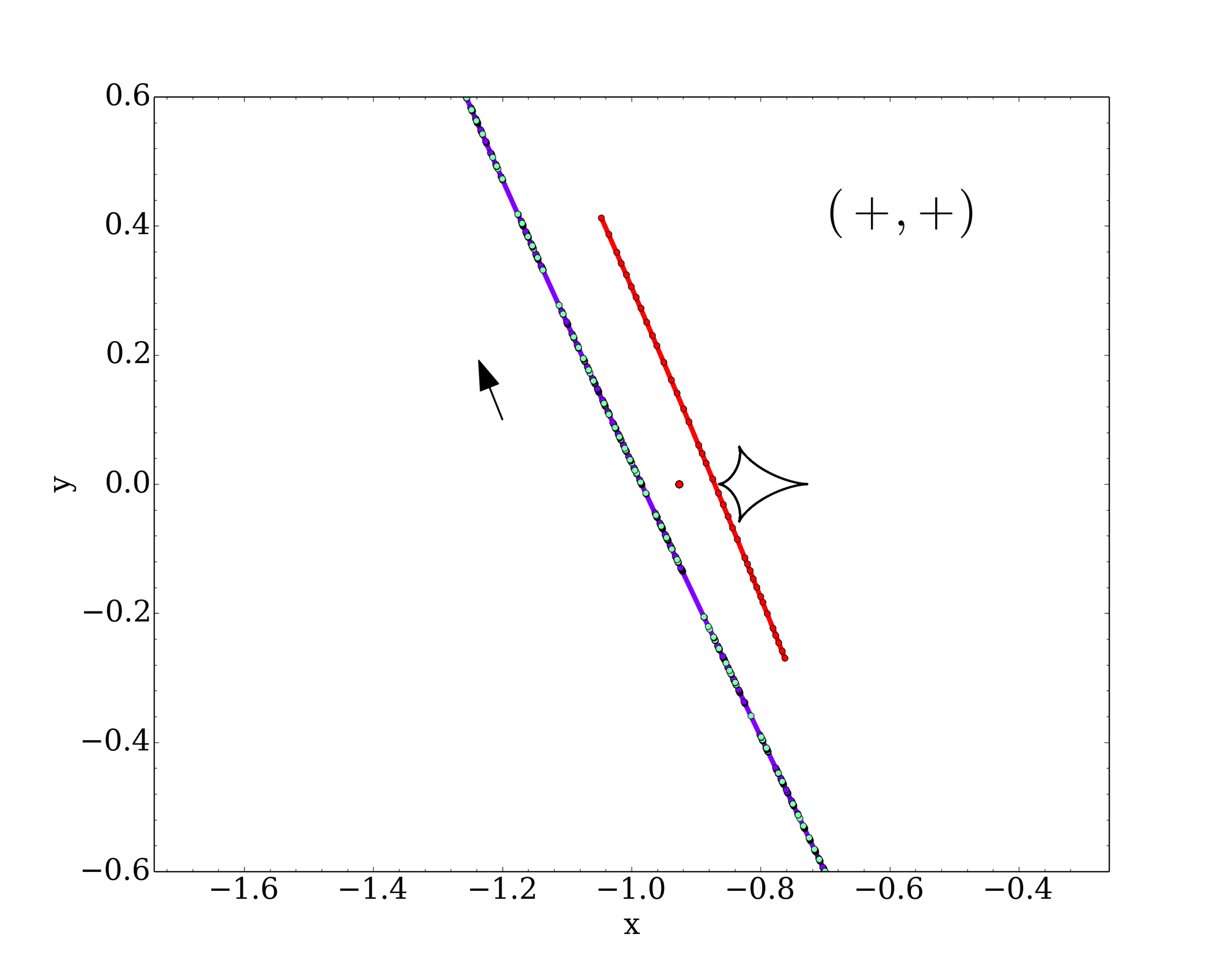}
\includegraphics[width=0.48\textwidth]{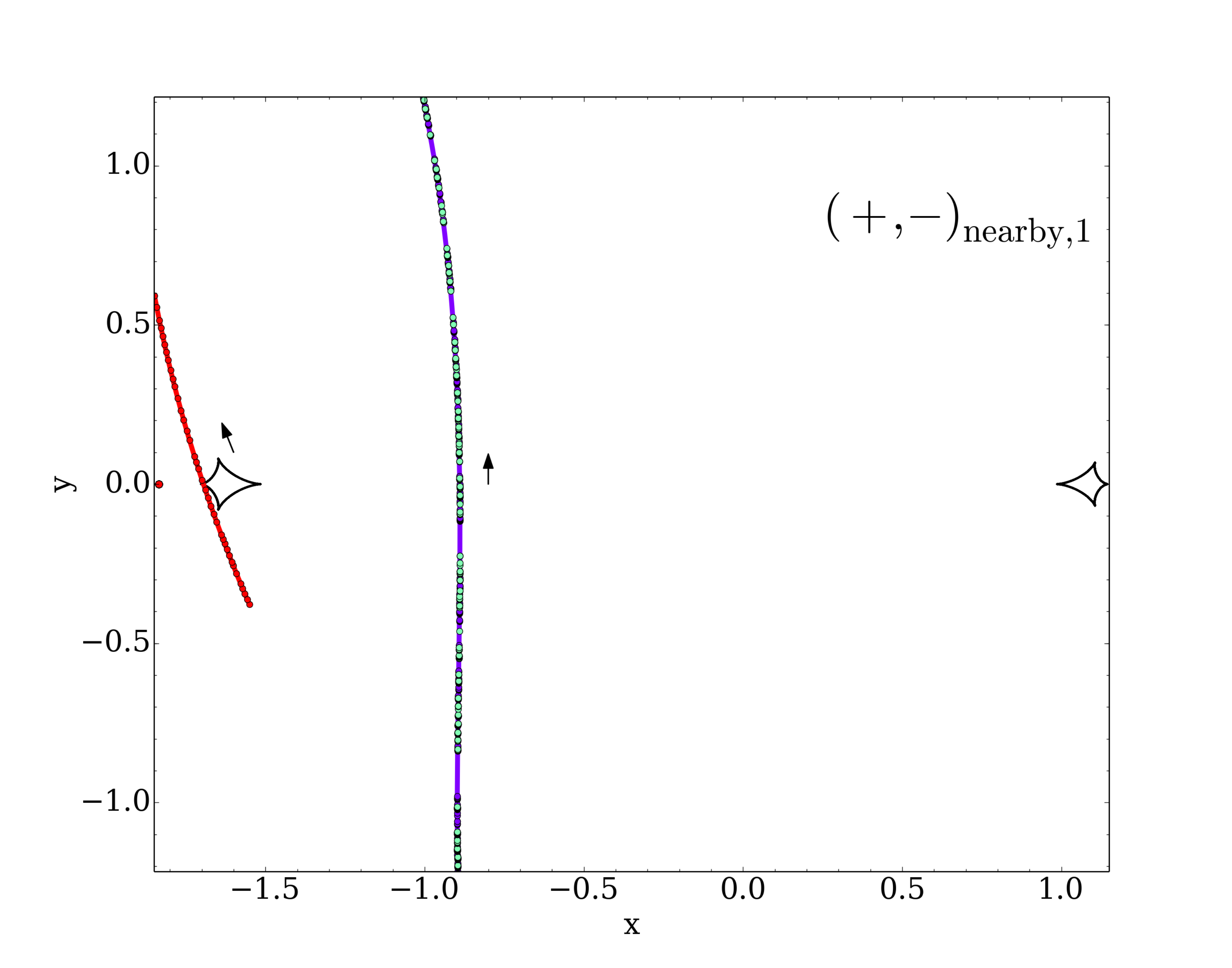}
\includegraphics[width=0.48\textwidth]{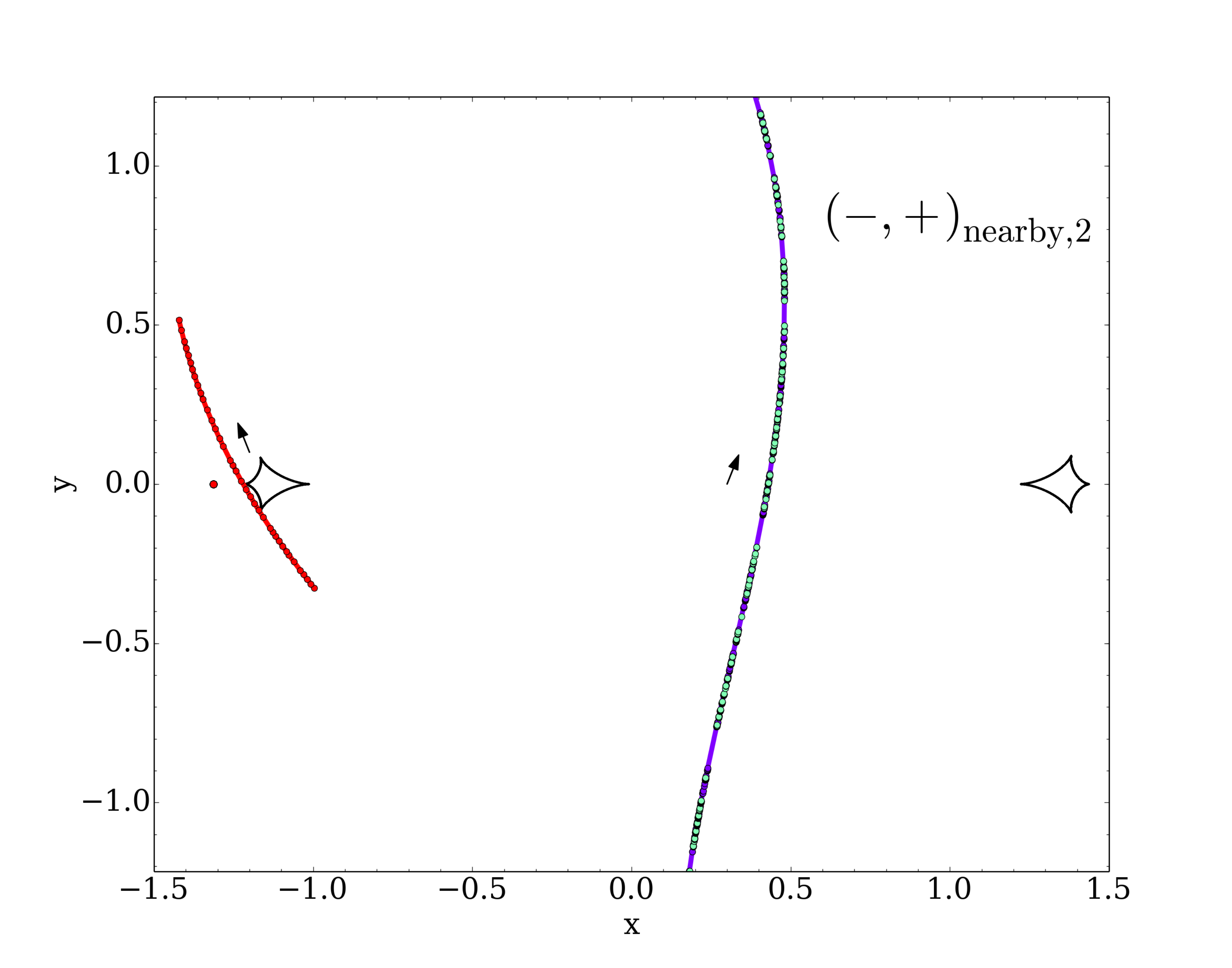}
\includegraphics[width=0.48\textwidth]{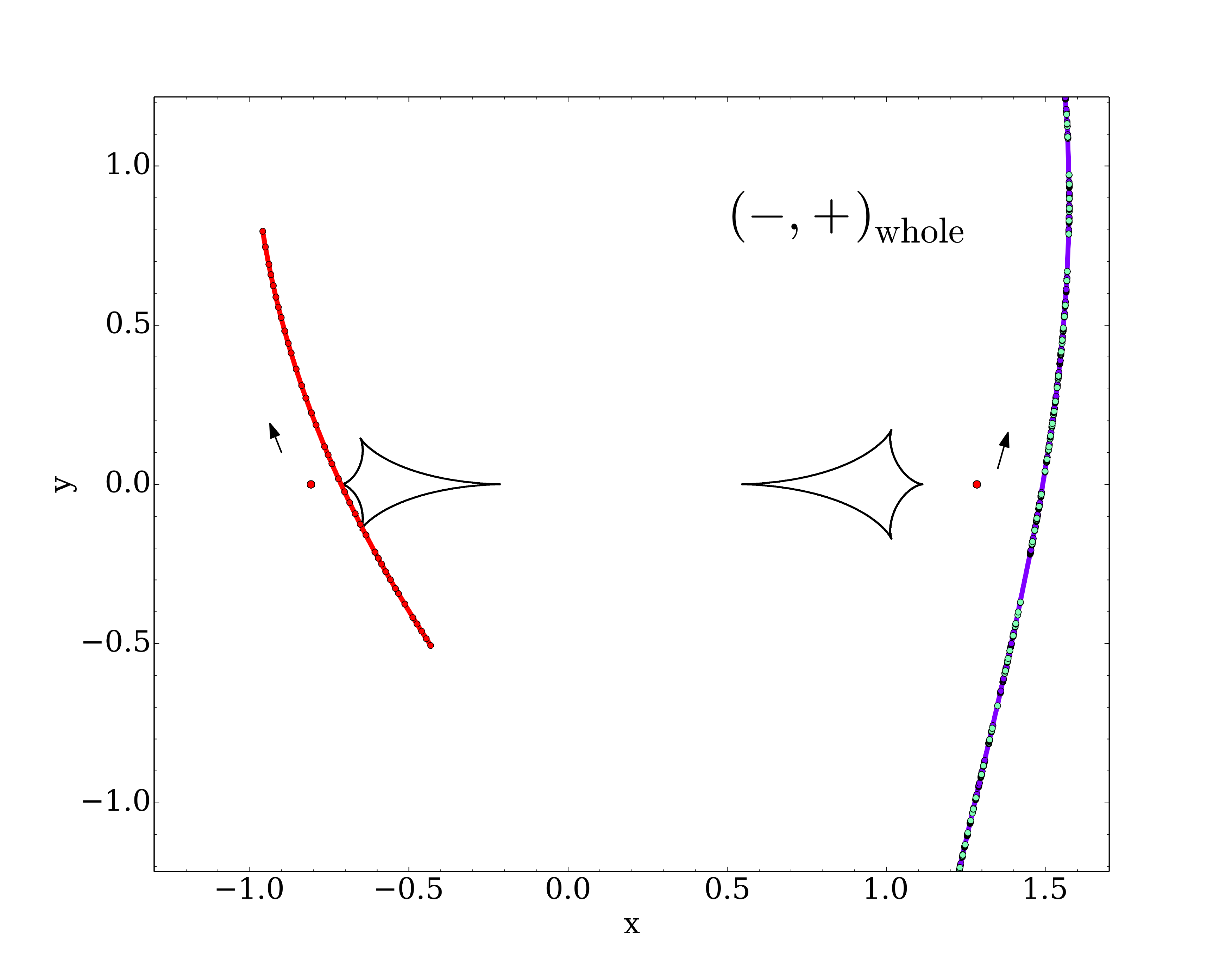}
\caption{Caustics and trajectories of the best-fit model and its degenerate solutions. The red (blue) curve shows the source
trajectory as seen from Spitzer (Earth). The red dots mark the positions of the lens components.}\label{cau}
\end{figure}

\indent\indent The best-fit models are the (+,+) and ($-$,$-$) solutions listed in Table~\ref{parms1}. After finding these two solutions, we looked for large-parallax degenerate solutions by setting the initial guess of parallax parameters to large values and running a longer MCMC. The other four-fold degenerate solutions are all found by this method. As discussed above, the source trajectories seen from Earth and the satellite in the``opposite'' solution could pass on either the opposite side of the whole lens system (two solutions) or of the nearby component (four solutions). Therefore, there are six possible large-parallax degenerate solutions. We present the  light curves and caustic plots for the (+, +), (+,$-$)$_\text{nearby,1}$, ($-$,+)$_\text{nearby,2}$ and ($-$, +)$_{\rm whole}$ in Figures~\ref{lc} and~\ref{cau}. The solutions of $u_0\pm$ degeneracy are similar to each other (the caustics are almost the same, with trajectories reflected about the $x$-axis), and we only present figures for one solution for each pair. Moreover, in event OGLE-2017-BLG-1130, the binary signal is only detected by {\it Spitzer}, and it is easier to see the difference between different solutions if the source trajectories seen by {\it Spitzer} are fixed on the caustics plots. Therefore, we choose to present figures of solutions with the same sign of $u_0$ as seen from {\it Spitzer} (and different signs of $u_0$ as seen from Earth).

The three pairs of large-parallax solutions represented in the three diagrams [(+,$-$)$_{\rm nearby,1}$, ($-$,+)$_{\rm nearby,2}$ and ($-$,+)$_{\rm whole}$] can qualitatively explain the data, but they are actually not viable. In addition to their larger $\chi^2$, these solutions all have excessive negative blending. The $F_S$ parameters measured from the OGLE data set are too large, implying that the unmagnified source fluxes, $I_{S, \rm OGLE}=18-2.5\log_{10}(F_{S,{\rm OGLE}})<17.3$\footnote{We use an I=18 flux scale in our fit, i.e., $I_{\rm base, OGLE}$=18 corresponds to 1 flux unit.}, are clearly ruled out by the total baseline of OGLE data, $I_{\rm s, OGLE}=18.69$.

\subsubsection{Close/Wide degeneracy}
\indent \indent Here we consider the ``close-wide'' degeneracy.
The best-fit model listed in Section~\ref{4fold} is the ``wide'' ($s>$1) solution, and we discuss the ``close'' ($s<$1) solution here. In this case, the four-fold degeneracy is reduced into the two-fold $u_0\pm$ degeneracy. First, for a close lens there is only one diamond-shaped caustic so there are only two large-parallax solutions. Second, the large-parallax solutions are disfavored by $\Delta \chi^2>100 $. Their parameters are shown in Table~\ref{parms2}. 
These two solutions have $\chi^2$ larger than the best-fit model by about 70 and hence are rejected.

\subsubsection{Other solutions}
It is possible to reproduce the two peaks in light curves seen in {\it Spitzer} data provided that the trajectories seen by {\it Spitzer} pass the diamond-shaped caustics
at different angles while the other parameters remain approximately the same. For example, a source trajectory at roughly 90 degrees to the one shown in Figure 3 would pass the
bottom cusp and then the right-most cusp, producing two bumps as seen in the light curve. There are two such solutions, one
corresponding to the best-fit model and the other 
corresponding to the ``close'' solution. These two solutions have $\chi^2$ larger than the best-fit model by more than 130 and hence are rejected.

We have also tried binary-source models. These fail by $\Delta\chi^2\sim300$ for {\it Spitzer} only and by $\Delta\chi^2\sim310$ for combined data sets, so they are not considered.  

\section{Physical Parameters}
\label{sec:phys}

The amplitude of the parallax vector, $\pi_\e=0.097\pm 0.005, $\footnote{We use the value in the $(-,-)$ solution hereafter because its $\chi^2$ is smaller. The $(+,+)$ solution has a similar microlens parallax amplitude $\pi_\e=0.095\pm 0.005$.} is well-measured.  Hence, if the Einstein radius $\theta_\e$ were also well measured, we could directly determine the lens mass $M=\theta_\e/\kappa\pi_\e$ and lens-source relative parallax, $\pi_{\rm rel}=\theta_\e\pi_\e$. Unfortunately, as is apparent from Table 1, the normalized source size, $\rho=\theta_\star/\theta_\e$, is barely detected. In fact, as we show below, $\rho$ is consistent with zero at the $\Delta\chi^2=1.5$
level.  The fact that $\rho$ is weakly constrained implies that $\theta_\e=\theta_\star/\rho$ is likewise weakly constrained. We will therefore ultimately require a Bayesian analysis to estimate the mass and distance of the lens system.

Even though $\rho$ is not strongly constrained, we must still measure $\theta_\star$ in order to make use of it at all.  This turns out to require a somewhat novel technique.

\subsection{Measurement of $\theta_\star$}
\label{sec:thetastar}

The usual path to measuring $\theta_\star$ \citep{Yoo:2004} is to start by measuring the source color and magnitude on an instrumental color-magnitude diagram, usually $(V-I, I)_S$, and to find the offset of this quantity from the clump, i.e., $\Delta(V-I, I) = (V-I, I)_S - (V-I, I)_{\rm cl}$. Then one determines the intrinsic position of the clump $(V-I, I)_{\rm cl,0}=(1.06,14.35)$ from the literature \citep{Nataf:2013,Bensby:2013}, and so $(V-I, I)_{S,0} = (V-I, I)_{\rm cl,0} + \Delta(V-I, I)$. Finally, one transforms from $V/I$ to $V/K$ using the color-color relations of \citet{bb88} and then applies the color/surface-brightness relations of \citet{Kervella:2004}.

In our case, unfortunately, we cannot measure $V_S$ because the $V$-band images from both OGLE and KMTNet are corrupted by diffraction spikes from a nearby bright, blue star.  Moreover, a frequently used back-up (for cases that $V$-band data are too poor to be used), namely an $H$-band light curve, is also not available in the present case.

We therefore introduce a novel approach to this problem by employing the {\it Spitzer} $3.6\,\mu$m (``$L$-band'') observations to determine the $(V-I)$ color.  In fact, there is a well-developed technology for contructing $VIL$ color-color relations for {\it Spitzer} microlensing data \citep{170event}. Normally, this is used when the {\it Spitzer} source flux is not well constrained by the microlensing light curve, which often occurs if the {\it Spitzer} data begin well after the peak.  In these cases, the well-measured $(V-I)_S$ color is then used to determine $(I-L)_S$ and thereby strongly constrain the {\it Spitzer} source flux (and therefore the magnification changes as a function of time).

\begin{figure*}
\centering
\centering
\includegraphics[width=0.48\textwidth]{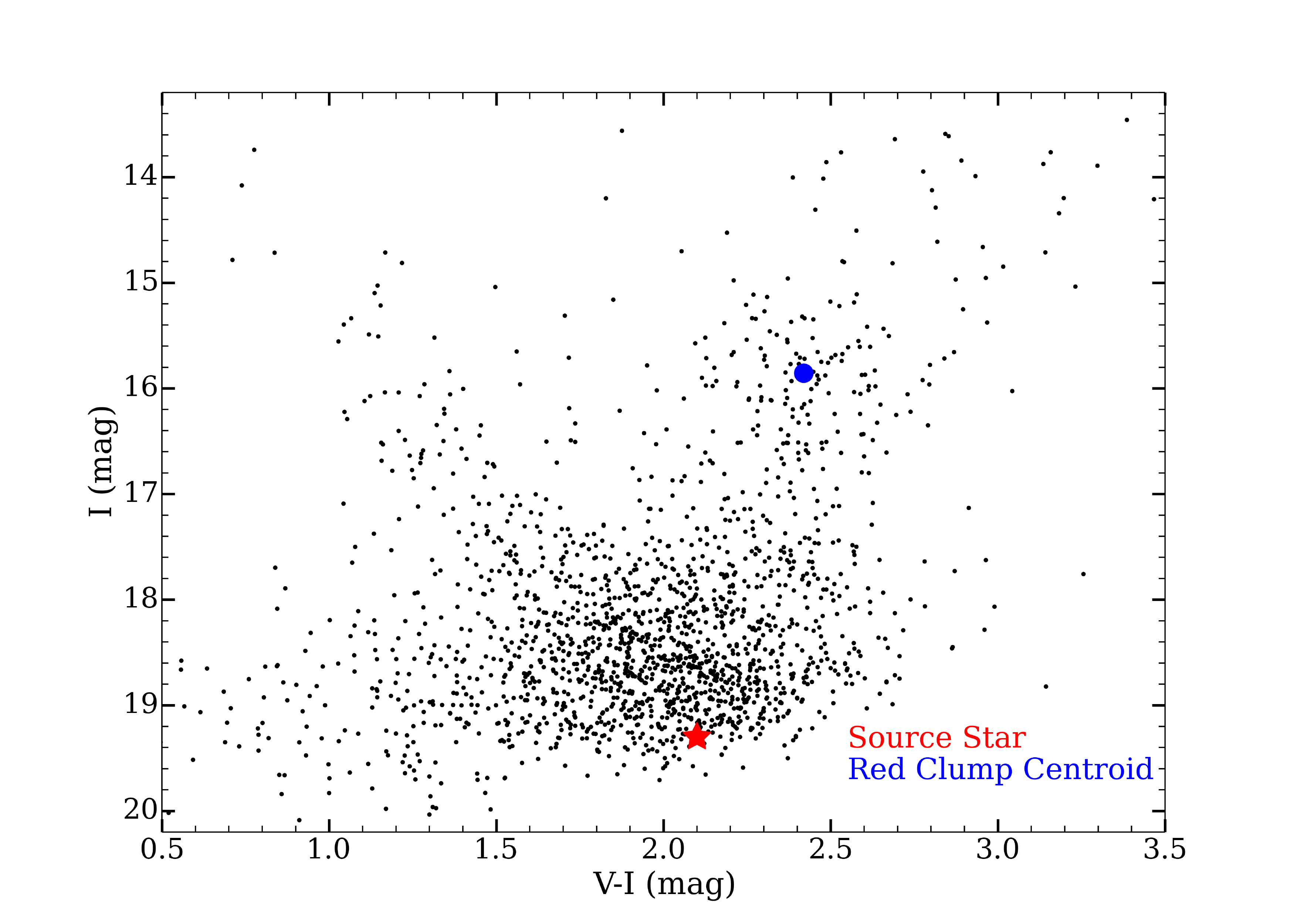}
\caption{OGLE-IV color magnitude diagram of the stars (black dots) within $2'\times2'$ of OGLE-2017-BLG-1130. The red and blue dots show the source star and the centroid of the red clump stars, respectively.} \label{fig:cmd}
\end{figure*}

In the present case, we invert this procedure.  From the measured $(I-L)_S=1.48\pm 0.15$ [or $(I-L)_S=1.44\pm 0.15$] color derived from the fits in Table~1, we find
$(V-I)_{S, \rm OGLE} = 2.10 \pm 0.15$ [or $(V-I)_{S, \rm OGLE} = 2.07 \pm 0.15$]. See Figure~\ref{fig:cmd}. Then, applying all the steps above, we find
\begin{equation}
\begin{split}
    \theta_\star &= 0.92\pm0.10~\muas \quad {\rm for~(+,+)~solution},\\
    \theta_\star &= 0.90\pm0.10~\muas \quad {\rm for~(-,-)~solution}.
\end{split}
\label{eqn:thetastar}
\end{equation}

\subsection{Bayesian Analysis}
\label{sec:bayes}
We begin our Bayesian analysis by extracting from the MCMC the best fit $a_{0,i}$ and covariance $c_{ij}$ of the three measured quantities $a_i=(\tilde v_{\rm l, hel}, \tilde v_{\rm b, hel}, t_{\rm E, hel})$. Here,
\begin{equation}
\bdv{\tilde{v}}_{\rm hel}=\bdv{\tilde{v}}_{\rm geo}+\bdv{v}_{\oplus,\perp};\qquad t_{\rm E,hel}=\frac{{\tilde v}_{\rm geo}}{{\tilde v}_{\rm hel}} t_{\rm E}
\end{equation}
are the helicentric velocity and timescale, where $v_{\oplus,\perp}(\rm N,E)=(-0.73, 27.22)\kms$, which is equivalent to $v_{\oplus,\perp}(\rm l,b)=(12.98, -23.94)\kms$.

We consider bulge sources and disk or bulge lenses drawn randomly from the Galactic model in \citet{Zhu:2017}, and for each trial we draw a mass of the primary star randomly from a \citet{Kroupa:2001} mass function.  We then calculate the resulting $\theta_{\rm E}=\sqrt{\kappa M\pi_{\rm rel}}$, ${\bdv{\tilde v}}_{\rm hel}={\bdv{\mu}}_{\rm hel}{\rm AU}/\pi_{\rm rel}$, $t_{\rm E, hel}=\theta_E/\mu_{\rm hel}$ and $\rho=\theta_\star/\theta_E$.

We then evaluate
\begin{equation}
\begin{split}
\chi^2_{\rm gal}&=\chi^2(\rho)+\chi^2_{\rm dyn},\\
\chi^2_{\rm dyn}&=\sum_{i,j=1}^3(a-a_0)_ib_{ij}(a-a_0)_j,
\end{split}
\end{equation}
where $a_i=(\tilde v_{\rm l, hel}, \tilde v_{\rm b, hel}, t_{\rm E, hel})$, $b\equiv c^{-1}$, and $\chi^2(\rho)$ represents the lower envelope of the $(\chi^2\ \rm vs.\ \rho)$ diagram derived from the MCMC \citep{Calchi Novati:2018}. We then weight all trials by the probability evaluated by combining $\chi^2_{\rm gal}$ and the microlensing rate contribution,
\begin{equation}
w_i=\exp(-\chi^2_{\rm gal,i}/2)\times\theta_{\rm E,i}\mu_{\rm i}.
\end{equation}

We also take into account the flux constraint on the lens. The blend flux is $I_b=19.9$ and $I_b=20.0$ for the $(+,+)$ and $(-,-)$ solutions. We find that the microlensed source is displaced from the ``baseline object'' by $0.22^{\prime\prime}\pm0.02^{\prime\prime}$. This implies that no more than about 50\% of the blended light could be due to the lens.  To be conservative, we set an upper limit of 75\%, which implies $I_l > 20.2$ and $I_l > 20.3$ in the two cases.  We then use these as the upper limits on the lens flux. We adopt the mass-luminosity relation
\begin{equation}
M_I = 4.4-8.5\log\biggl(\frac{M_{\rm prim}}{M_\odot}\biggr),
\end{equation}
where $M_I$ is the absolute magnitude in $I$-band and $M_{\rm prim}$ is the mass of the primary. Then the lens distance should satisfy
\begin{equation}
M_I+5\log\biggl(\frac{D_L}{10{\ \rm pc}}\biggr)+A_I\geq I_b,
\end{equation}
where $D_L$ is the distance to the lens and the extinction $A_I=I_{\rm RC}-I_{\rm RC,0}=1.52$. We reject trials that violate this relation.

\begin{figure*}
\centering
\includegraphics[width=0.9\textwidth]{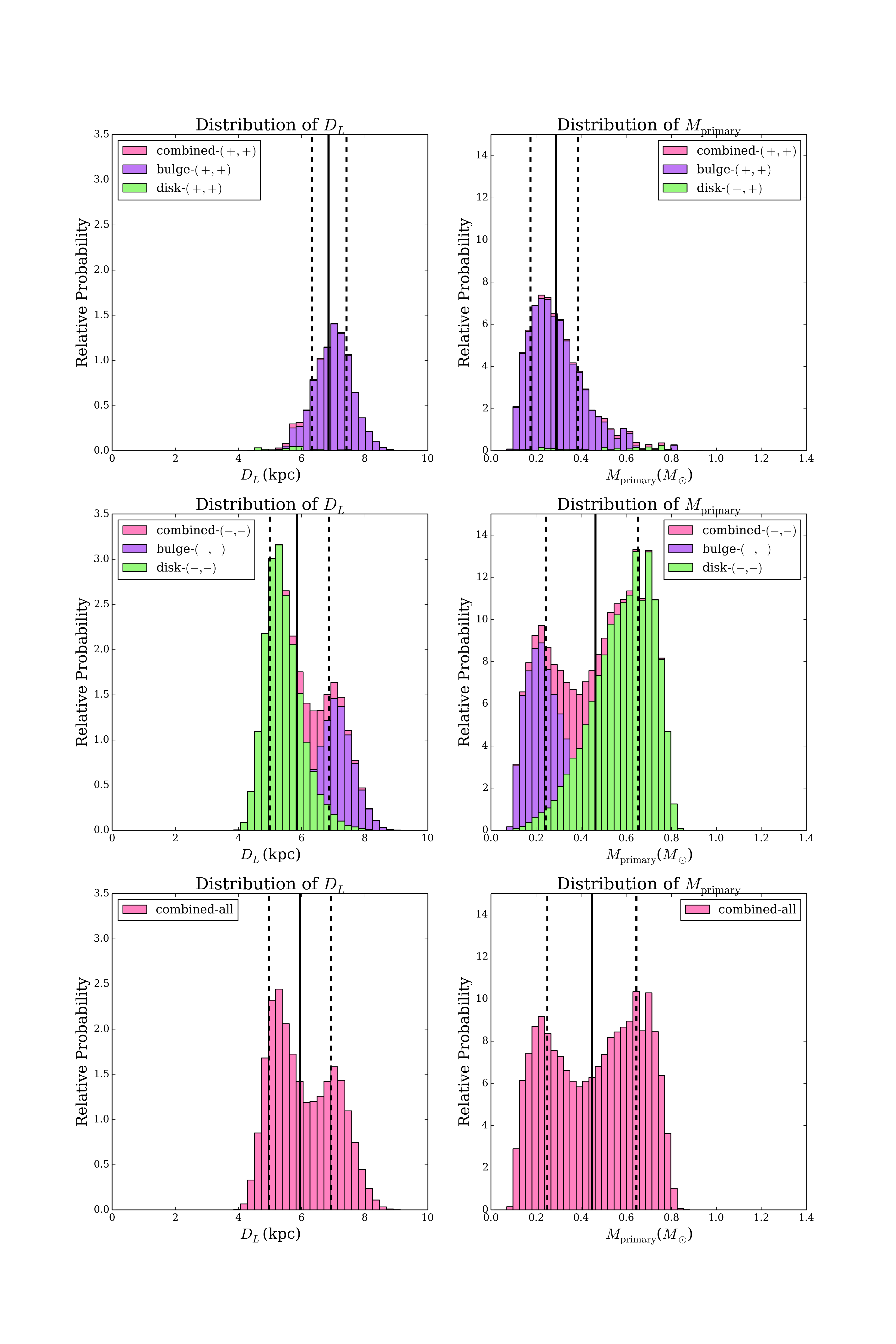}
\caption{Distribution of $D_L$ and $M_{\rm prim}$ from the Bayesian analysis. Top: the $(+,+)$ solution. Middle: the $(-,-)$ solution. Bottom: combined distributions. The distributions are arbitrarily normalized.}\label{fig:bayes}
\end{figure*}

The results of the Bayesian analysis are shown in Figure~\ref{fig:bayes}. For bulge lenses, the $(+,+)$ and $(-,-)$ solutions yield similar distributions of physical parameters. On the other hand, for disk lenses, the $(-,-)$ solution is strongly favored because its direction is right in the direction of Galactic rotation. The ratio between the probability of bulge and disk lenses in the $(-,-)$ solution is about 1/2.2, while the disk lens part of the $(+,+)$ solution is almost ruled out. In principle, the two solutions $(+,+)$ and $(-,-)$ should be weighted by $e^{-\chi^2/2}$. However, the difference in $\chi^2$ is well-within the margin of what can be produced by typical microlensing systematics. Hence, we just weight them by total probability and so obtain $M_{\rm prim}=0.45\pm0.20M_\odot$ and $D_L=5.9\pm1.0\, \kpc$.

\subsection{Future Resolution}
From 
\begin{equation}
M=\frac{\mu_\hel t_{\e,\hel}}{\kappa\pi_\e}
\end{equation}
and 
\begin{equation}
\pi_\rel=\pi_\e\mu_\hel t_{\e,\hel},
\end{equation}
we can measure the lens mass and lens-source relative parallax if a future determination of the lens-source relative heliocentric proper motion $\bdv{\mu}_\hel$ is  available. Because the errors in $\pi_\e$ and $t_\e$ are about 10\% and 5\%, the mass and relative parallax can ultimately be constrained to $\pm15\%$, provided that the proper-motion measurement is more precise than this.

The vector proper motion measurement would also decisively rule out (or possibly confirm one of) the other solutions that we analyzed in Section~\ref{minima}. As discussed above, the larger parallax solutions are extremely unlikely to be correct due to their large $\chi^2$ and excessive negative blending. The proper motion measurement can confirm this conclusion.

To assess when such a measurement can first be made, we first estimate the expected proper motion as a function of the lens mass $M=(1+q)M_{\rm prim}$ and quantities that are directly measured from the light curve
\begin{equation}\label{eqn:bmuhel}
\begin{split}
\bmu_\hel &= {\pi_\rel\over\au}\bdv{\tilde{v}}_{\rm hel}\\
&= {\kappa\pi_\e^2 M\over\au}(\bdv{\tilde{v}}_{\rm geo}+\bdv{v}_{\oplus,\perp})\\
&= \kappa M\biggl({\bpi_\e\over t_\e} +
   {\pi_\e^2\over\au}\bdv{v}_{\oplus,\perp}\biggr)
\end{split}
\end{equation}
For the two cases, this yields
\begin{equation}\label{eqn:bmuhel}
\begin{split}
\bmu_{\hel_{+,+}}(\rm N,E) &= (-4.7,+3.5)\masyr \biggl({M\over M_\odot} \biggr);\\
\bmu_{\hel_{-,-}}(\rm N,E) &= (+5.2,+2.9)\masyr \biggl(\frac{M}{M_\odot}\biggr),
\end{split}
\end{equation}
i.e., similar amplitudes $\mu_\hel \simeq 5.9\,\masyr (M/M_\odot)$.

Based on the experience of \citet{ob05169bat}, who resolved the equally-bright source and lens of OGLE-2005-BLG-169 at a separation of $\sim60~$mas,
we can see that such a measurement using present-day instrumentation would require a 15 year wait for an $M\sim 0.7\,M_\odot$ ($M_{\rm prim}\sim 0.48\,M_\odot$) lens.  From Figure~\ref{fig:bayes} and Equation~\ref{eqn:bmuhel}, this would imply only a 50\% probability of separately resolving the source and lens.  However, by this time it is very likely that next generation (``30 meter'') telescopes with adaptive optics will be operating.  Since these will have roughly three times better resolution than the current 8-10m telescopes, the lens and source can almost certainly be resolved at first light of these instruments.

\section{Discussion}\label{sec:dis}
\indent \indent We analyzed the binary-lensing event OGLE-2017-BLG-1130 in which the binary anomaly was only detected by the {\it Spitzer} Space Telescope. We found the lens parameters by fitting the space-based data, and we measured the microlensing parallax using ground-based observations. 

This event provides strong evidence that some binary signals (as predicted by \citealt{mao:1991}) can be missed by observations from the ground alone but detected by {\it Spitzer}, especially for wide and close binaries. Although space-based data are normally used to measure the microlensing parallax, it is possible that some interesting signals, for example, planetary signals, can only be seen from {\it Spitzer}. In event OGLE-2014-BLG-0124, the planetary signal was independently detected from {\it Spitzer}, and if the trajectories had been slightly different, the planetary signal could have been detected by {\it Spitzer} and missed from the ground. In addition, such binaries may affect the observed event timescale distributions \citep{Wegg:2017, Mroz:2017}, as in event OGLE-2017-BLG-1130 the timescale fitted from ground data is about 10 days shorter than the real case. Therefore, the role that {\it Spitzer} plays in microlensing observations is more than functioning as a parallax satellite, and it will produce more results of scientific interest in the future.

The binary-lensing event OGLE-2017-BLG-1130 is peculiar in another aspect. We show that the normal four-fold space-based degeneracy can in principle become eight-fold: $(+,-)\times$(same, opposite nearby component \& close to primary, opposite nearby component \& close to secondary, opposite whole binary). This eight-fold degeneracy should not occur frequently because as it requires at least three conditions: (1) the mass ratio is close to unity because the timescale set by one component of the binary should be similar to the timescale set by the other, (2) the source trajectory is nearly normal to the binary-lens axis and, (3) the binary separation is sufficiently large. 

\acknowledgements
This work has been supported in part by the National Natural Science Foundation of China (NSFC) grants 11333003 and 11390372 (SM). 
This research has made use of the KMTNet system operated by the Korea Astronomy and Space Science Institute (KASI) and the data were obtained at three host sites of CTIO in Chile, SAAO in South Africa, and SSO in Australia. 
The OGLE project has received funding from the National Science Centre, Poland, grant MAESTRO 2014/14/A/ST9/00121 to AU. 
Work by WZ, YKJ, and AG were supported by AST-1516842 from the US NSF.
WZ, IGS, and AG were supported by JPL grant 1500811.
Work by C.H. was supported by the grant (2017R1A4A101517) of
National Research Foundation of Korea.
Work by YS was supported by an appointment to the NASA Postdoctoral Program at the Jet Propulsion Laboratory,
California Institute of Technology, administered by Universities Space Research Association
through a contract with NASA.
This work is based (in part) on observations made with the Spitzer
Space Telescope, which is operated by the Jet Propulsion Laboratory,
California Institute of Technology under a contract with NASA.
Support for this work was provided by NASA through an award issued by
JPL/Caltech.

\clearpage
\begin{table*}[!h]
\centering
\caption{Best Solutions with 8-fold Degeneracy (the ``wide'' solutions).}\label{parms1}
\begin{tabular}{ccccccccc}
\hline
					&($+$,$+$)			&($-$,$-$)			&$(+,-)_\text{nearby,1}$&$(-,+)_\text{nearby,1}$		&(+,$-$)$_\text{nearby,2}$		&($-$,+)$_\text{nearby,2}$		&(+,$-$)$_\text{whole}$		&($-$,+)$_\text{whole}$ \\
\hline
$\chi^2/DOF$			&2866.4/2866			&2864.3/2866			&2919.1/2866			&2933.9/2866			&2983.5/2866			&2971.1/2866			&2962.9/2866			&2994.5/2866\\
$t_0{({\rm HJD}')}$		&7931.33$\pm$1.93	&7931.46$\pm$1.75	&7948.31$\pm$0.57	&7947.80$\pm$0.72	&7945.42$\pm$1.07	&7945.72$\pm$0.92	&7943.40$\pm$0.99	&7943.72$\pm$0.17\\
$u_0$	        			&0.881$\pm$0.052		&-0.907$\pm$0.046	&0.890$\pm$0.095		&-0.969$\pm$0.118	&0.435$\pm$0.051		&-0.424$\pm$0.051	&1.517$\pm$0.008		&-1.470$\pm$0.013\\
$t_E\text{(day)}$		&49.76$\pm$2.37		&49.39$\pm$2.33		&37.28$\pm$0.51		&39.73$\pm$0.47		&43.28$\pm$1.33		&39.64$\pm$0.86		&29.62$\pm$1.12		&26.51$\pm$0.60\\
$\rho$				&0.005$\pm$0.002		&0.006$\pm$0.002		&0.012$\pm$0.001		&0.011$\pm$0.001		&0.001$\pm$0.001		&0.001$\pm$0.001		&0.015$\pm$0.003		&0.007$\pm$0.001\\
$\pi_{\rm E,N}$		&-0.079$\pm$0.006	&0.088$\pm$0.006		&0.605$\pm$0.037		&-0.584$\pm$0.046	&-1.194$\pm$0.030	&1.208$\pm$0.034		&-1.558$\pm$0.023	&1.616$\pm$0.015\\
$\pi_{\rm E,E}$		&0.052$\pm$0.003		&0.041$\pm$0.002		&0.016$\pm$0.010		&0.088$\pm$0.011		&-0.083$\pm$0.048	&-0.234$\pm$0.035	&-0.055$\pm$0.031	&-0.271$\pm$0.006\\
$\alpha\text{(deg)}$	&115.67$\pm$1.50		&-115.01$\pm$1.44	&90.86$\pm$0.97		&-91.31$\pm$1.26		&-81.16$\pm$2.14		&81.02$\pm$1.52		&-81.58$\pm$1.20		&80.29$\pm$0.16\\
$s$					&2.95$\pm$0.05		&2.98$\pm$0.07		&3.06$\pm$0.02		&3.02$\pm0.03$		&2.81$\pm$0.02		&2.87$\pm$0.02		&1.98$\pm$0.04		&2.09$\pm$0.02\\
$q$					&0.447$\pm$0.037		&0.456$\pm$0.031		&1.49$\pm$0.13		&1.72$\pm$0.16		&0.801$\pm$0.04		&0.842$\pm$0.048		&0.432$\pm$0.025		&0.629$\pm$0.020\\
I-L 					&1.48$\pm$0.15		&1.44$\pm$0.15		&-0.31$\pm$0.18		&-0.19$\pm$0.16		&-0.38$\pm$0.10		&-0.54$\pm$0.10		&0.21$\pm$0.06		&0.28$\pm$0.06\\
$F_{\rm S,OGLE}$		&0.30				&0.31				&2.89				&2.84				&2.80				&3.08				&2.15				&2.09\\
$F_{\rm B,OGLE}$		&0.17				&0.16				&-2.40				&-2.35				&-2.31				&-2.58				&-1.65				&-1.46\\
\hline
\end{tabular}

\end{table*}

\begin{table}
\centering
\caption{The ``close'' solutions with $u_0\pm$ Degeneracy}\label{parms2}
\begin{tabular}{ccc}
\hline
			&(+,+)				&($-$, $-$)					\\
\hline
$\chi^2/DOF$	&2933.7/2866			&2933.4/2866			\\
$t_0{({\rm HJD}')}$		&7949.43$\pm$0.08	&7949.45$\pm$0.08	\\
$u_0$		&0.132$\pm$0.012		&-0.128$\pm$0.018	\\
$t_E\text{(day)}$		&50.07$\pm$4.29		&51.46$\pm$3.70		\\
$\rho$		&0.007$\pm$0.004		&0.007$\pm$0.004		\\
$\pi_{\rm E,N}$	&-0.063$\pm$0.006	&0.069$\pm$0.010		\\
$\pi_{\rm E,E}$	&0.059$\pm$0.006		&0.050$\pm$0.006		\\
$\alpha\text{(deg)}$		&113.84$\pm$1.31		&-113.93$\pm$1.19	\\
$s$			&0.393$\pm$0.011		&0.387$\pm$0.020		\\
$q$			&0.280$\pm$0.034		&0.275$\pm$0.037		\\
I-L 			&0.031				&0.029				\\
$F_{\rm S,OGLE}$	&0.232			&0.223\\
$F_{\rm B,OGLE}$	&0.264			&0.272	\\
\hline
\end{tabular}

\end{table}

\clearpage

\end{document}

%% file: author.tex
\author{Tianshu~Wang}
\affil{Physics Department and Tsinghua Centre for Astrophysics, Tsinghua University, Beijing 100084, China}

\author{S. Calchi Novati}
\affil{IPAC, Mail Code 100-22, Caltech, 1200 E. California Blvd., Pasadena, CA 91125, USA}
\affil{Spitzer Team}

\author{A. Udalski}
\affil{Warsaw University Observatory, Al. Ujazdowskie 4,00-478 Warszawa, Poland}
\affil{OGLE collaboration}

\author{A. Gould}
\affil{Spitzer Team}
\affil{Korea Astronomy and Space Science Institute, Daejon 34055, Korea}
\affil{Department of Astronomy, Ohio State University, 140 W. 18th Ave., Columbus, OH 43210, USA}
\affil{Max-Planck-Institute for Astronomy, K\"{o}nigstuhl 17, 69117 Heidelberg, Germany}
\affil{KMTnet}

\author{Shude~Mao}
\affil{Physics Department and Tsinghua Centre for Astrophysics, Tsinghua University, Beijing 100084, China}
\affil{National Astronomical Observatories, Chinese Academy of Sciences, 20A Datun Road, Chaoyang District, Beijing 100012, China}
\affil{Jodrell Bank Centre for Astrophysics, School of Physics and Astronomy, The University of Manchester, Oxford Road, Manchester M13 9PL, UK}

\author{W.~Zang}
\affil{Physics Department and Tsinghua Centre for Astrophysics, Tsinghua University, Beijing 100084, China}
\affil{Department of Physics, Zhejiang University, Hangzhou, 310058, China}
\nocollaboration

\correspondingauthor{Tianshu~Wang}
\email{wts15@mails.tsinghua.edu.cn}

\author{C.~Beichman}
\affil{NASA Exoplanet Science Institute, MS 100-22, California Institute of Technology, Pasadena, CA 91125, USA}
\author{G.~Bryden}
\affil{Jet Propulsion Laboratory, California Institute of Technology, 4800 Oak Grove Drive, Pasadena, CA 91109, USA}
\author{S.~Carey}
\affil{Spitzer Science Center, MS 220-6, California Institute of Technology, Pasadena, CA, US}
\author{B.~S.~Gaudi}
\affil{Department of Astronomy, Ohio State University, 140 W. 18th Ave., Columbus, OH 43210, USA}
\author{C.~B.~Henderson}
\affil{Jet Propulsion Laboratory, California Institute of Technology, 4800 Oak Grove Drive, Pasadena, CA 91109, USA}
\author{Y.~Shvartzvald}
\affil{Jet Propulsion Laboratory, California Institute of Technology, 4800 Oak Grove Drive, Pasadena, CA 91109, USA}
\affil{NASA Postdoctoral Program Fellow}
\author{J. C. Yee}
\affil{Harvard-Smithsonian Center for Astrophysics, 60 Garden St., Cambridge, MA 02138, USA}
\affil{KMTnet}

\collaboration{(Spitzer Team)}
\author{P. Mr\'{o}z}
\affil{Warsaw University Observatory, Al. Ujazdowskie 4,00-478 Warszawa, Poland}

\author{R. Poleski}
\affil{Warsaw University Observatory, Al. Ujazdowskie 4,00-478 Warszawa, Poland}
\affil{Department of Astronomy, Ohio State University, 140 W.18th Ave., Columbus, OH 43210, USA}

\author{J. Skowron}
\affil{Warsaw University Observatory, Al. Ujazdowskie 4, 00-478 Warszawa, Poland}

\author{M. K. Szyma\'{n}ski}
\affil{Warsaw University Observatory, Al. Ujazdowskie 4,00-478 Warszawa, Poland}

\author{I. Soszy\'{n}ski}
\affil{Warsaw University Observatory, Al. Ujazdowskie 4,00-478 Warszawa, Poland}

\author{S. Koz{\l}owski}
\affil{Warsaw University Observatory, Al. Ujazdowskie 4,00-478 Warszawa, Poland}

\author{P. Pietrukowicz}
\affil{Warsaw University Observatory, Al. Ujazdowskie 4,00-478 Warszawa, Poland}

\author{K. Ulaczyk}
\affil{Warsaw University Observatory, Al. Ujazdowskie 4,00-478 Warszawa, Poland}

\author{M. Pawlak}
\affil{Warsaw University Observatory, Al. Ujazdowskie 4,00-478 Warszawa, Poland}

\collaboration{(OGLE Collaboration)}

\author{M. D. Albrow}
\affil{University of Canterbury, Department of Physics and Astronomy, Private Bag 4800, Christchurch 8020, New Zealand}

\author{S.-J. Chung}
\affil{Korea Astronomy and Space Science Institute, Daejon 34055, Korea}
\affil{Korea University of Science and Technology, Daejeon 34113, Korea}

\author{C. Han}
\affil{Department of Physics, Chungbuk National University, Cheongju 28644, Republic of Korea}

\author{K.-H. Hwang}
\affil{Korea Astronomy and Space Science Institute, Daejon 34055, Korea}

\author{Y. K. Jung}
\affil{Harvard-Smithsonian Center for Astrophysics, 60 Garden St., Cambridge, MA 02138, USA}

\author{Y.-H. Ryu}
\affil{Korea Astronomy and Space Science Institute, Daejon 34055, Korea}

\author{I.-G. Shin}
\affil{Harvard-Smithsonian Center for Astrophysics, 60 Garden St., Cambridge, MA 02138, USA}


\author{W. Zhu}
\affil{Canadian Institute for Theoretical Astrophysics, University of Toronto, 60 St George Street, Toronto, ON M5S 3H8, Canada}
\affil{Spitzer Team}

\author{S.-M. Cha}
\affil{Korea Astronomy and Space Science Institute, Daejon 34055, Korea}
\affil{School of Space Research, Kyung Hee University, Yongin, Kyeonggi 17104, Korea}

\author{D.-J. Kim}
\affil{Korea Astronomy and Space Science Institute, Daejon 34055, Korea}

\author{H.-W. Kim}
\affil{Korea Astronomy and Space Science Institute, Daejon 34055, Korea}

\author{S.-L. Kim}
\affil{Korea Astronomy and Space Science Institute, Daejon 34055, Korea}
\affil{Korea University of Science and Technology, Daejeon 34113, Korea}

\author{C.-U. Lee}
\affil{Korea Astronomy and Space Science Institute, Daejon 34055, Korea}
\affil{Korea University of Science and Technology, Daejeon 34113, Korea}

\author{D.-J. Lee} 
\affil{Korea Astronomy and Space Science Institute, Daejon 34055, Korea}

\author{Y. Lee}
\affil{Korea Astronomy and Space Science Institute, Daejon 34055, Korea}
\affil{School of Space Research, Kyung Hee University, Yongin, Kyeonggi 17104, Korea}

\author{B.-G. Park}
\affil{Korea Astronomy and Space Science Institute, Daejon 34055, Korea}
\affil{Korea University of Science and Technology, Daejeon 34113, Korea}

\author{R. W. Pogge}
\affil{Department of Astronomy, Ohio State University, 140 W. 18th Ave., Columbus, OH 43210, USA}
\collaboration{(KMTNet Collaboration)}